\documentclass[aps, prl, superscriptaddress, reprint, nobibnotes]{revtex4-2}
\usepackage{amsmath,amssymb}
\usepackage{color}
\usepackage{comment}
\usepackage{bm,graphicx,hyperref}
\hypersetup{%
  breaklinks = {true},
  citecolor = {blue},
  colorlinks = {true},
  linkcolor = {red},}

\usepackage{lineno}

\begin{document}

\title{Advantages of non-Hookean coupling in a measurement-fueled two-oscillator engine}

\author{A. Rodin}
\affiliation{Yale-NUS College, 16 College Avenue West, 138527, Singapore}
\affiliation{Centre for Advanced 2D Materials, National University of Singapore, 117546, Singapore}
\affiliation{Department of Materials Science and Engineering, National University of Singapore, 117575, Singapore}

\begin{abstract}

A quantum engine composed of two oscillators with a non-Hookean coupling is proposed.
Unlike the more common quantum heat engines, the setup introduced here does not require heat baths as the energy for the operation originates from measurements.
The engine follows the coupling $\rightarrow$ measurement $\rightarrow$ decoupling $\rightarrow$ extraction cycle.
Using a Gaussian term as a prototypical non-harmonic interaction, it is shown that the fueling process facilitates the decoupling step.
Numerical simulations are used to demonstrate the measurement-driven fueling, as well as the reduced decoupling energy. 

\end{abstract}	
\maketitle
\emph{Introduction.} Drawing inspiration from their classical counterparts, quantum engines often employ similar cycles as the traditional heat engines, but with the role of the working fluid, typically performed by a gas, played by quantum components.~\citep{Bhattacharjee2021}
For example, the quantum Otto cycle can use two-level systems~\citep{Huang2012, Thomas2018}, one~\citep{DelCampo2014, Rossnagel2014, Kosloff2017, Cavaliere2022, Fei2022} or multiple~\citep{Boubakour2023} harmonic oscillators, or photons~\citep{Zhang2014} as the working fluid.
Notably, despite their quantum nature, these engines need two heat baths to operate, just like their classical versions.

Recently, a novel type of quantum engine was introduced, requiring a single bath.~\citep{Elouard2017, Elouard2017a, Elouard2018, Das2019}
In this setup, the energy does not originate from the hot reservoir but from projective measurements.~\citep{Bresque2021, Manikandan2022, Jussiau2023}
In the protocol described in Ref.~\citep{Jussiau2023}, an entanglement-based engine operates along the following cycle.
One starts with the system components (e.g., qubits or harmonic oscillators) in their respective ground states, isolated from each other.
Next, an interaction between the components is switched on, and the composite system relaxes to its ground state by transferring the excess energy to the cold bath.
After that, projective measurements on each component bring them to their local eigenstates, after which the interaction is switched off.
Finally, because the components will, generally, not be in their ground states after the measurement, their excitation energy can be used to perform work, returning the subsystems to their ground states and restarting the cycle.

One of the implementations of this elegant engine described in Ref.~\citep{Jussiau2023} is based on two harmonic oscillators coupled by a tunable Hooke interaction so that the system Hamiltonian is $H_0 + k(x_1 - x_2)^2/2$, where $H_0$ is the independent-oscillator component and $x_{1/2}$ are the oscillator displacements.
Combining $kx_1^2 + kx_2^2$ with $H_0$ produces a ``local" Hamiltonian, leaving the cross term as the coupling.~\citep{Jussiau2023}
A clear advantage of the linear interaction is evident: if the two oscillators are in their eigenstates, the expectation value of this term is zero due to its odd parity in the position coordinates.
Consequently, switching the interaction on (when both oscillators are at their ground state) or switching it off (after the projective measurement) has no energy cost.

There also exists a downside associated with the harmonic coupling.
Adding (removing) the interaction amounts to setting the force constant $0\rightarrow k$ ($k\rightarrow 0$), meaning that, in addition to controlling the coupling, this process modifies the local Hamiltonian.
In other words, when the interaction is introduced (removed), the oscillators undergo an effective compression (expansion).
A way to mitigate this effect is by including a ``counter term" $k(x_1^2+x_2^2)/2$ to ensure that the local Hamiltonian remains constant, which adds complexity to the system.
Alternatively, one could accept the modification of the local Hamiltonian during the switching process.
As a consequence, switching on the coupling would require energy input to squeeze the oscillators, while switching it off would drain a part of the energy because reduced confinement decreases the spacing between the oscillator levels, reducing the efficiency of the engine.

The purpose of this work is to propose a new type of bath-free engine driven by entanglement.
The principle of operation closely follows the two-harmonic oscillator implementation of the engine described in Ref.~\citep{Jussiau2023} with two key differences: the absence of the cold bath and the non-Hookean interaction between the oscillators.
The bath is removed since this engine's operation does not rely on it being at the entangled ground state.
The reason behind using the non-Hookean coupling is twofold.
First, a more general interaction opens the door to a wider variety of implementation approaches.
More importantly, however, this coupling type can make the interaction-switching process substantially simpler.
The discussion below introduces the formalism behind this engine and then describes two implementation schemes, supported by numerical calculations.

\emph{Model.} The engine is composed of two harmonic oscillators with frequencies $\Omega_1$ and $\Omega_2$, coupled by a position-dependent interaction term.
It is convenient to express all energies in terms of the first oscillator's energy $\hbar\Omega_1$ and the lengths using its quantum oscillator length $l_1$ so that the general Hamiltonian for the system becomes

\begin{equation}
    \hat{H} 
    = \left(\hat{a}^\dagger \hat{a} + \frac{1}{2}\right)
    + \omega\left(\hat{b}^\dagger \hat{b} + \frac{1}{2}\right)
    + \Phi(\hat{x}_1,\hat{x}_2)\,,
    \label{eqn:Hamiltonian}
\end{equation}
where $\omega = \Omega_2 / \Omega_1$ and $\Phi(\hat{x}_1,\hat{x}_2)$ is the interaction.

A natural way to describe the system is using the Fock basis $|j\rangle_1\otimes|k\rangle_2$, where $j$ ($k$) is the energy level of the first (second) oscillator.
Writing $|j\rangle_1\otimes|k\rangle_2$ as $|j,k\rangle$ for brevity, the Hamiltonian matrix elements in this basis are

\begin{align}
    \langle u, v|\hat{H}|j,k\rangle
    =&
    \left[\left(j + \frac{1}{2}\right)
    + \omega\left(k + \frac{1}{2}\right)\right]\delta_{ju}\delta_{kv}
    \nonumber
    \\
    +&
    \langle u, v| \Phi(\hat{x}_1,\hat{x}_2)|j,k\rangle
    \label{eqn:H_matrix_element}
\end{align}
with

\begin{align}
    &  \langle u, v| \Phi(\hat{x}_1,\hat{x}_2)|j,k\rangle
    \nonumber
    \\
     =
     &
     \int dx_1dx_2
      \Phi(x_1,x_2)
     \nonumber
     \\
     \times &
     \Psi_{1,j}(x_1)\Psi_{2,k}(x_2)
      \Psi_{1,u}^*(x_1)\Psi_{2,v}^*(x_2)\,,
      \label{eqn:Phi_Matrix}
\end{align}
where $\Psi_{g,h}(x) = \langle x|h\rangle_g$ are the harmonic oscillator wavefunctions.

As discussed above, the system starts in the state $|0,0\rangle$ with the interaction switched off.
Introducing an attractive coupling between the modes lowers the system's energy by $|\langle 0, 0|\Phi(\hat{x}_1,\hat{x}_2)|0,0\rangle|$, making it an energetically-favorable process.
Performing the projective measurement takes the system to state $|j,k\rangle$ with energy $(j+1/2) + \omega(k+1/2)+\langle j, k|\Phi(\hat{x}_1,\hat{x}_2)|j,k\rangle$.
Crucially, if $\Phi$ decreases with $x_1$ and $x_2$, because the wavefunctions for $j>0$ and $k>0$ are wider than the ground state ones, the interaction energy $\langle j, k|\Phi(\hat{x}_1,\hat{x}_2)|j,k\rangle$ will not be greater in magnitude than $\langle 0, 0|\Phi(\hat{x}_1,\hat{x}_2)|0,0\rangle$.
Thus, turning off the interaction would require less energy than the energy released by introducing the coupling.
In fact, the higher the excitation level, the smaller the required separation energy is. 
After the energy extraction, the oscillators return to $|0,0\rangle$, and the cycle repeats.

Two general configurations for the two-oscillator system are explored numerically below.
To perform numerical calculations, the highest energy level for each oscillator is set to 50, resulting in $51^2 = 2,601$ basis states for the composite system and a Hamiltonian matrix with $2,601^2 = 6,765,201$ elements.
All computations are performed using the {\scshape julia} programming language.~\citep{Bezanson2017}
The plots are made using Makie.jl package~\citep{Danisch2021} using the color scheme designed for colorblind readers.~\citep{Wong2011}
The scripts used for computing and plotting can be found at https://github.com/rodin-physics/quantum-oscillator-engine.

\emph{Parallel oscillators.} The conceptually simplest engine configuration involves two parallel oscillators.
In this case, the interaction between them can be written as $\Phi(x_1 - x_2)$.
Although it is, in principle, feasible to compute the matrix elements in Eq.~\eqref{eqn:Phi_Matrix} directly for a particular $\Phi$, the two-dimensional integral can cause issues in numerical calculations. 
Fortunately, it is possible to reduce the integral to a one-dimensional form.
By writing the interaction term as the Fourier transform

\begin{equation}
    \Phi(x) 
    =
    \int_{-\infty}^\infty d\gamma
    \underbrace{\left[\int_{-\infty}^\infty \frac{dz}{2\pi}\, \Phi(z) e^{i\gamma z}\right]}_{\Phi_\gamma}\, e^{-i\gamma x}\,,
    \label{eqn:Phi_FT_2Osc}
\end{equation}
one can split the $x_1$ and $x_2$ integrals in Eq.~\eqref{eqn:Phi_Matrix}:

\begin{align}
    & \langle u, v|\Phi(\hat{x}_1-\hat{x}_2)|j,k\rangle
    \nonumber
    \\
    =&
      \int d\gamma\,\Phi_\gamma 
    \left[\int dx_1 \Psi_{1,u}^*(x_1) \Psi_{1,j}(x_1)
    e^{-i\gamma x_1}\right]
    \nonumber
    \\
    \times &
    \left[\int dx_2 \Psi_{2,v}^*(x_2)\Psi_{2,k}(x_2)
    e^{i\gamma x_2}\right]\,.
      \label{eqn:Separated_Variables_2Osc}
\end{align}

From the brackets in Eq.~\eqref{eqn:Separated_Variables_2Osc}, it is useful to define

\begin{align}
     \Xi\left(\gamma  l, u,j\right)&=
     \int dx\, \Psi_{u}^*(x) \Psi_{j}(x)
    e^{-i\gamma x}
    \nonumber
    \\
     &=\int  \frac{dx}{l}
    \frac{e^{-\frac{ x^2}{l^2}}
    e^{-i\gamma x}}{\sqrt{2^uu!}\sqrt{2^jj!}\sqrt{\pi}}
    H_u\left(\frac{x}{l}\right)
    H_j\left(\frac{x}{l}\right)
    \nonumber
    \\
    &= e^{-\frac{\gamma^2l^2}{4}}\int dy\,
    H_u\left(y\right)
    H_j\left(y\right)
    \frac{
    e^{-\left(y+i\frac{\gamma l}{2}\right)^2}}{\sqrt{\pi2^{u+j}u!j!}}
    \nonumber
    \\
    &= 
      e^{-\frac{\gamma^2l^2}{4}}
    \sqrt{\frac{j!}{u!}}
    \left(\frac{\gamma l}{i\sqrt{2}}\right)^{u-j}L_j^{u-j}\left(\frac{\gamma^2l^2}{2}\right)\,,
    \label{eqn:Xi_def}
\end{align}
where $l$ is the quantum harmonic oscillator length, $H_n$ are the Hermite polynomials, and $L_m^{n - m}$ are the generalized Laguerre polynomials.
The integral in Eq.~\eqref{eqn:Xi_def} was obtained from a table of integrals,~\citep{Gradshteyn} leading to

\begin{align}
    & \langle u, v|\Phi(\hat{x}_1-\hat{x}_2)|j,k\rangle
    \nonumber
    \\
    =&
      \int d\gamma\,\Phi_\gamma 
    \Xi\left(\gamma, u,j\right) 
    \Xi\left(-\gamma  \lambda, v, k\right)\,,
      \label{eqn:Interaction_Element_2Osc}
\end{align}
where $\lambda = l_2 / l_1$.

When performing numerical calculations, it is best to precompute the $\Phi$-matrix for a particular interaction as these integrals are the slowest step given the large size of the Hamiltonian matrix.
The two-oscillator Hamiltonian, on the other hand, is diagonal in the Fock basis, requiring no additional computation.
Expressing the time in terms of the oscillator periods $t = 2\pi \tau / \Omega_1$, one can use the precomputed $\Phi$ matrix to obtain $\hat{H}$ in the two-oscillator basis and get the time evolution operator

\begin{equation}
    \hat{\mathcal{U}}(\tau - \tau') = \exp\left[-2\pi i \hat{H}(\tau-\tau')\right]\,.
    \label{eqn:Time_Evolution}
\end{equation}

For the purposes of illustration, it is easiest to work with identical oscillators so that $\lambda = \omega = 1$.
The interaction term $\Phi$ is set to $\Phi_0 e^{-x^2/2\sigma^2}$ with $\Phi_\gamma =\Phi_0\sigma e^{-\gamma^2\sigma^2 / 2}/\sqrt{2\pi}$.
The choice of the $\Phi$ is guided by convenience: one can easily control the magnitude and the extent of the interaction for a Gaussian expression, allowing it to serve as a prototypical coupling.

\begin{figure}
    \centering
    \includegraphics[width = \columnwidth]{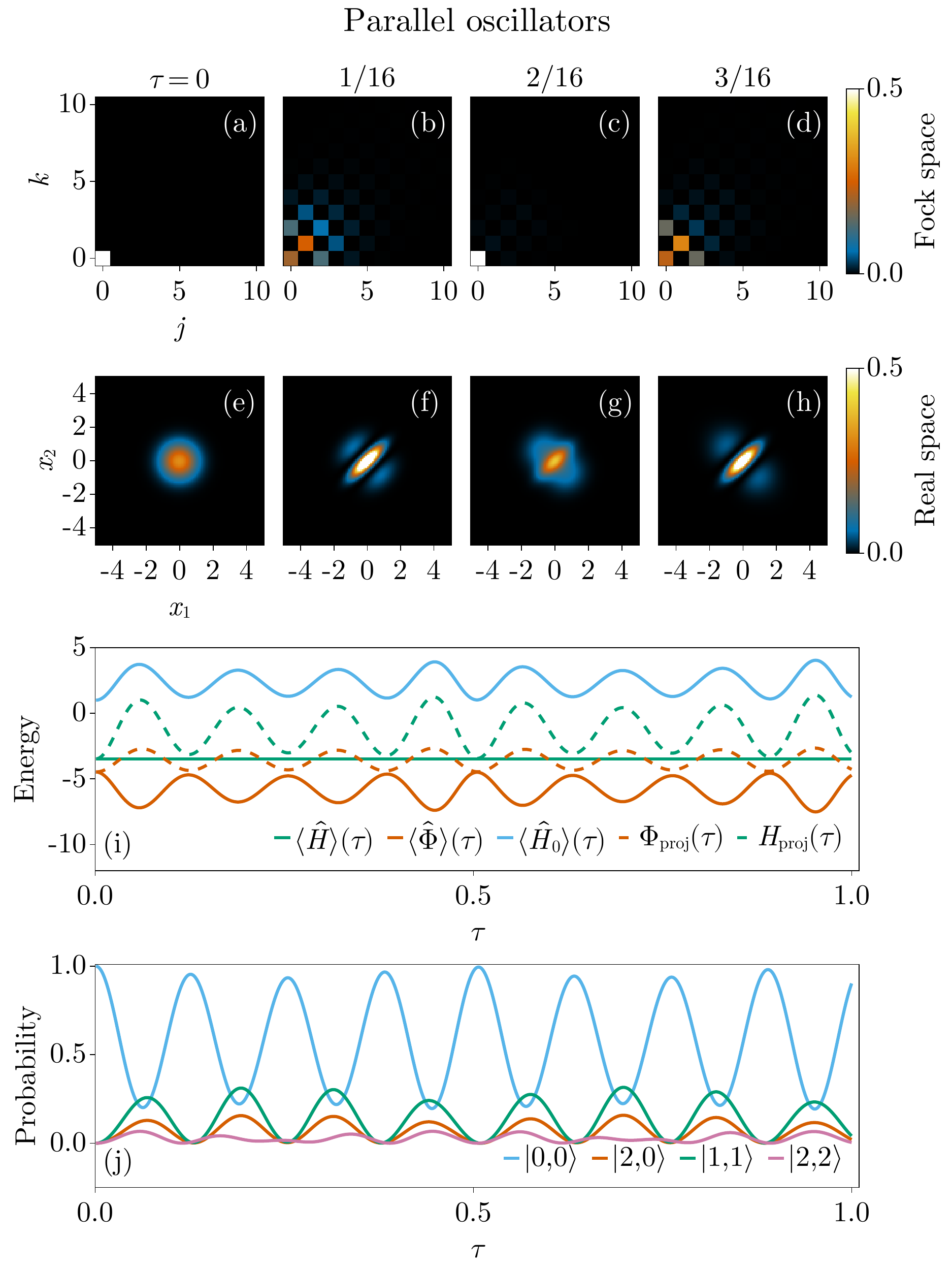}
    \caption{\emph{Evolution of coupled parallel oscillators.} Two identical parallel oscillators, initialized at their ground state, interact via a Gaussian coupling term with amplitude $\Phi_0 = -10$ and width $\sigma = 1/2$. Panels (a)-(d) show the probability distribution of the system in the Fock space; (e)-(h) is the real-space representation of the probability distribution. For clarity, only the lowest 11 Fock states are shown for each oscillator out of 51 used in the calculation.
    Panel (i) shows the system's energies. Solid lines are the expectation values for the interaction, free oscillator, and total energies as the system undergoes unitary evolution. The latter is equal to the sum of the two former ones and is constant in time. Dashed lines show the average interaction and total energies following a projective measurement performed at time $\tau$.
    Panel (j) gives the time-dependent probability of several lowest-energy independent-oscillator states. The reduced occupation of the ground state $|0,0\rangle$ coincides with the weakened post-projection interaction energy.}
    \label{fig:2Osc}
\end{figure}

As an example, consider a system with $\Phi_0 = -10$ and $\sigma = 1/2$.
At $\tau = 0$, when the two oscillators are in their respective ground states, the interaction is switched on.
To evolve the state, $\hat{\mathcal{U}}(\delta\tau)$ for $\delta\tau = 0.001$ is calculated and then applied repeatedly on the initial state.
Figure~\ref{fig:2Osc}(a)-(h) shows the subsequent time evolution of the two-oscillator wavefunction, both in Fock and real space.
One can see that, as the occupation of the Fock states varies in time, only the states where the parity of the wavefunctions is the same for the two oscillators are permitted.
The real-space wavefunction shows a positive correlation between the positions of the two oscillators, as expected from an attractive interaction.
The finite probability occupation of the non-$|0,0\rangle$ states indicates that projective measurements will be able to take the system to one of the excited states.

Introducing the coupling sets the system energy to $1 - 2\sqrt{5}$, where the first term comes from the ground state energy and the second one is the interaction, as can be computed from $\langle 0, 0|\Phi|0,0\rangle$.
Because $\hat{H}$ commutes with $\hat{\mathcal{U}}$, the expectation value for the total energy $\langle 0,0|\hat{\mathcal{U}}^\dagger \hat{H}\hat{\mathcal{U}}|0,0\rangle = 1 - 2\sqrt{5}$ is constant, as confirmed by numerical results in Fig.~\ref{fig:2Osc}(i).
On the other hand, the expectation value of the interaction energy, given by $\langle 0,0|\hat{\mathcal{U}}^\dagger \hat{\Phi}\hat{\mathcal{U}}|0,0\rangle$, as well as the expectation energy of the $\hat{H}_0$ oscillate in time with their sum adding to $\langle \hat{H}\rangle$.

Measuring the system takes it to one of $|j,k\rangle$ states.
Because $\hat{H}_0$ is diagonal in the oscillator eigenstates, the mean post-measurement oscillator energy is identical to $\langle \hat{H}_0\rangle(\tau)$.
To compute the mean interaction energy after the projection, one uses $\sum_{j,k} \langle j,k|\hat{\Phi}|j,k\rangle \Psi^2_{jk}$.
In agreement with the discussion above, Fig.~\ref{fig:2Osc}(i) confirms that the mean post-projection interaction energy is shallower than $\langle 0,0|\hat{\Phi}|0,0\rangle$.

The reduced post-measurement interaction energy indicates that the system is likely to end up in one of the excited states.
To validate this point, the probability of several lowest-energy states as a function of time is plotted in Fig.~\ref{fig:2Osc}(j).
One can see that the time periods with the shallowest projected interaction energy correspond to instances where the probability of the ground state is the lowest.
In fact, measuring the system at these moments will produce an excited state $\approx 80\%$ of the time for the system parameters chosen here.

\emph{Perpendicular oscillators.} The second way of setting up the system is to have the oscillators move in perpendicular directions.
For this configuration, one can view the setup as either two independent oscillators interacting via $\Phi(|\mathbf{x}_1 - \mathbf{x}_2|)$ with $\hat{x}_1 \perp \hat{x}_2$ or as a single two-dimensional oscillator with a controllable central potential $\Phi$.
Either way, the interaction term takes the form $\Phi(x_1^2 + x_2^2)$.
It is important to note that for the perpendicular arrangement, the mode coupling is only made possible by the non-Hookean form of $\Phi$.
Otherwise, $\Phi(|\mathbf{x}_1 - \mathbf{x}_2|) = k(x_1^2+x_2^2) / 2$ simply changes the confining harmonic potential without actually mixing the perpendicular modes.
Although the two-oscillator setup is, perhaps, more interesting since it involves intrinsic system interactions, there are certain advantages to the central-potential configuration also.
The ability to control the coupling by adding or removing an external potential offers a way of exploring the system without interfering with its internal dynamics.

To keep as much similarity with the discussion above as possible, the interaction is chosen to be $\Phi_0 e^{-(x_1^2 + x_2^2) / 2\sigma^2}$.
Because of its Gaussian form, it is possible to evaluate the matrix element directly from Eq.~\eqref{eqn:Phi_Matrix}:

\begin{align}
    & \langle u, v| \Phi(x_1^2+ x_2^2) |j,k\rangle
    \nonumber
    \\
    =&
    \Phi_0 
    \left[\int dx_1 \Psi_{1,u}^*(x_1) \Psi_{1,j}(x_1)
    e^{- x_1^2/2\sigma^2}\right]
    \nonumber
    \\
    \times &
    \left[\int dx_2 \Psi_{2,v}^*(x_2)\Psi_{2,k}(x_2)
    e^{- x_2^2/2\sigma^2}\right]
    \nonumber
    \\
    =&
    \Phi_0 
    \Sigma(1/2\sigma^2 , u, j)
    \Sigma(\lambda^2/2\sigma^2 , v, k)\,.
\end{align}
where

\begin{align}
   \Sigma(\alpha l^2, u, j)
   &=
   \int dx\, \Psi_{u}^*(x) \Psi_{j}(x)
    e^{-\alpha x^2}
    \nonumber
    \\
    &=\int  \frac{dx}{l}
    \frac{e^{-\frac{ x^2}{l^2}}
    e^{-\alpha x^2}}{\sqrt{2^uu!}\sqrt{2^jj!}\sqrt{\pi}}
    H_u\left(\frac{x}{l}\right)
    H_j\left(\frac{x}{l}\right)
    \nonumber
    \\
    &= \int dy\,
    H_u\left(y\right)
    H_j\left(y\right)
    \frac{ e^{-y^2(1+\alpha l^2)}}{\sqrt{2^uu!}\sqrt{2^jj!}\sqrt{\pi}}
    \nonumber
    \\
    &=
     \frac{(u+j)!}{\left(\frac{u+j}{2}\right)!\sqrt{u!j!}}\left(\frac{1}{1+\alpha l^2}\right)^{\frac{u + j + 1}{2}}
     \left(-\frac{\alpha l^2}{2}\right)^{\frac{u+j}{2}}
     \nonumber
     \\
     &\times 
     F\left(-u,-j;\frac{1-u-j}{2};\frac{1+\alpha l^2}{2\alpha l^2}\right)
    \,,
    \label{eqn:Sigma}
\end{align}
if $u+j$ is an even number and 0 otherwise.
Here, $F$ is the Gaussian hypergeometric function.

\begin{figure}
    \centering
    \includegraphics[width = \columnwidth]{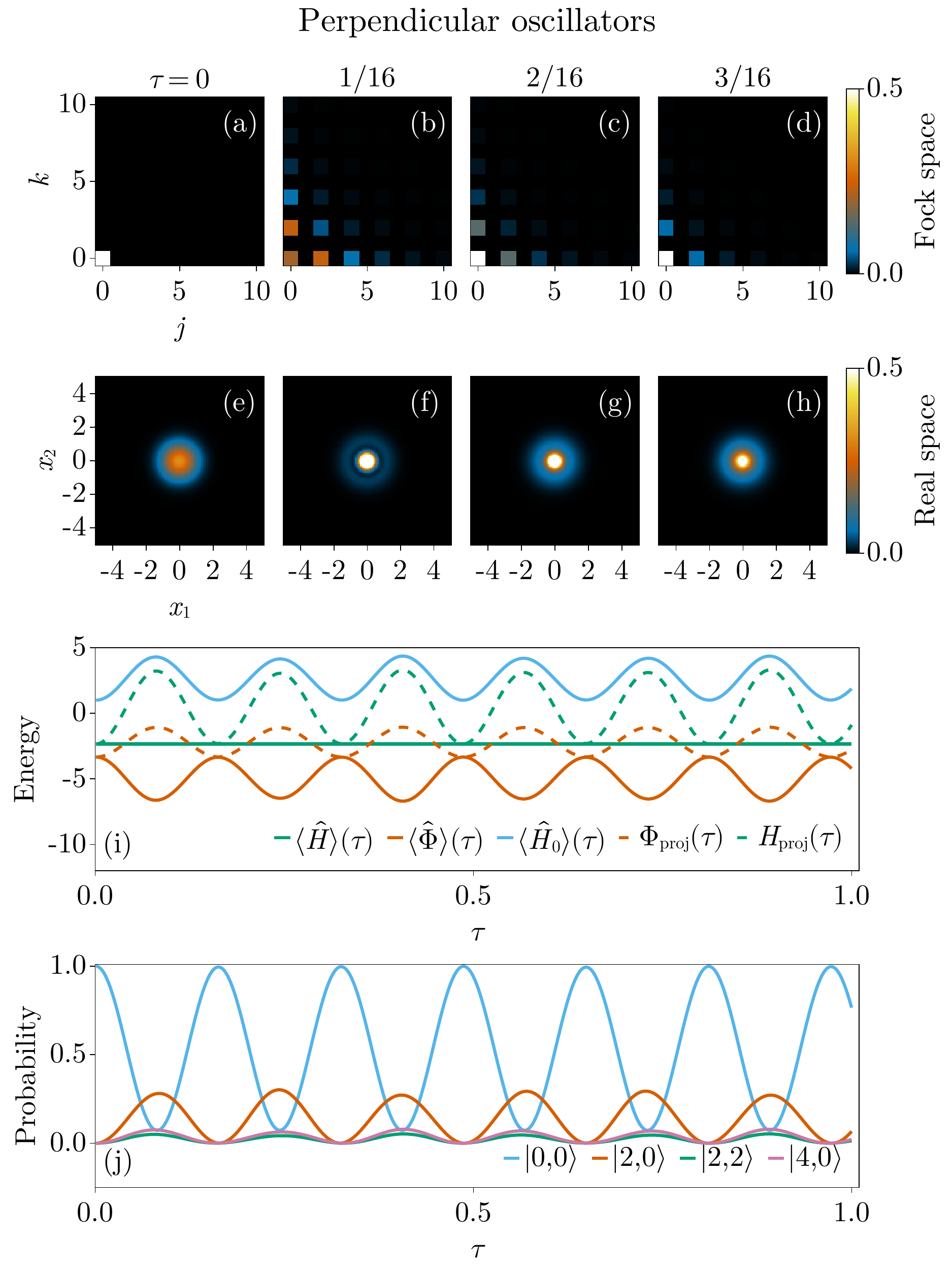}
    \caption{\emph{Evolution of coupled perpendicular oscillators.} Two identical perpendicular oscillators, initialized at their ground state, interact via a Gaussian coupling term with amplitude $\Phi_0 = -10$ and width $\sigma = 1/2$. Panels (a)-(d) show the probability distribution of the system in the Fock space; (e)-(h) is the real-space representation of the probability distribution. For clarity, only the lowest 11 Fock states are shown for each oscillator out of 51 used in the calculation.
    Because of the interaction symmetry, fewer excited Fock states are allowed compared to the parallel configuration.
    Panels (i) and (j) follow the same format as Fig.~\ref{fig:2Osc}(i) and (j).}
    \label{fig:2D}
\end{figure}

Similarly to the parallel oscillators, the time-dependent wavefunction in Fig.~\ref{fig:2D}(a)-(h) shows how the occupation of the excited energy levels changes in time.
Because of the symmetry imposed by $\Phi$, there is more restriction on which excited states are allowed when the system is initialized in $|0,0\rangle$: only in states composed of even single-oscillator states are permitted.
The shape of the two-oscillator wavefunction remains azimuthally symmetric due to the interaction form.
As above, one can see in Fig.~\ref{fig:2D}(i) that the total energy remains constant, while the interaction and free-oscillator energies vary in time.
Performing a measurement increases the total energy, as expected.

Figure~\ref{fig:2D}(j) shows the time dependence of the level occupation for several of the lowest energy levels.
Just as for the parallel setup, the lowest occupation of the $|0,0\rangle$ level corresponds with the shallowest post-projection interaction energy.
One can see that, for this configuration, the minimum occupation of the ground state is almost zero so that the measurement is essentially guaranteed to produce an excited state.

\emph{Summary and outlook.} This work has demonstrated the advantage of non-Hookean coupling in measurement-powered two-oscillator engines.
The operation of the engine requires one to switch the interaction between the oscillators on and off.
It was shown that, due to this interaction profile, the energy cost of turning off the coupling is reduced when the system is taken to an excited level by a projective measurement.
Thus, in addition to fueling the engine, the measurement also prepares the system to be decoupled by lowering the energy cost of the interaction switching.
In addition to making the switching more favorable energetically, this type of interaction also allows a greater variety of implementation schemes not possible for a harmonic coupling.
In particular, the engine can be composed of perpendicular oscillator modes, realized by either two independent oscillators or a single two-dimensional oscillator with the role of the interaction played by an external central potential.

There are several ways to build on the results presented here.
First, it is worth exploring the extension of this setup to include more oscillators.
In addition, by working with different interaction profiles, such as as (screened) Couloumb, one could search for ways to optimize the energetics of the switching process while maximizing the efficiency.
Finally, a protocol involving repeated measurements in order to ``pump" the system to a higher energy level to further reduce the coupling while providing an enhanced energy output could be investigated.

The author acknowledges the National Research Foundation, Prime Minister Office, Singapore, under its Medium Sized Centre Programme and the support by Yale-NUS College (through Start-up Grant).


\begin{thebibliography}{21}%
\makeatletter
\providecommand \@ifxundefined [1]{%
 \@ifx{#1\undefined}
}%
\providecommand \@ifnum [1]{%
 \ifnum #1\expandafter \@firstoftwo
 \else \expandafter \@secondoftwo
 \fi
}%
\providecommand \@ifx [1]{%
 \ifx #1\expandafter \@firstoftwo
 \else \expandafter \@secondoftwo
 \fi
}%
\providecommand \natexlab [1]{#1}%
\providecommand \enquote  [1]{``#1''}%
\providecommand \bibnamefont  [1]{#1}%
\providecommand \bibfnamefont [1]{#1}%
\providecommand \citenamefont [1]{#1}%
\providecommand \href@noop [0]{\@secondoftwo}%
\providecommand \href [0]{\begingroup \@sanitize@url \@href}%
\providecommand \@href[1]{\@@startlink{#1}\@@href}%
\providecommand \@@href[1]{\endgroup#1\@@endlink}%
\providecommand \@sanitize@url [0]{\catcode `\\12\catcode `\$12\catcode `\&12\catcode `\#12\catcode `\^12\catcode `\_12\catcode `\%12\relax}%
\providecommand \@@startlink[1]{}%
\providecommand \@@endlink[0]{}%
\providecommand \url  [0]{\begingroup\@sanitize@url \@url }%
\providecommand \@url [1]{\endgroup\@href {#1}{\urlprefix }}%
\providecommand \urlprefix  [0]{URL }%
\providecommand \Eprint [0]{\href }%
\providecommand \doibase [0]{https://doi.org/}%
\providecommand \selectlanguage [0]{\@gobble}%
\providecommand \bibinfo  [0]{\@secondoftwo}%
\providecommand \bibfield  [0]{\@secondoftwo}%
\providecommand \translation [1]{[#1]}%
\providecommand \BibitemOpen [0]{}%
\providecommand \bibitemStop [0]{}%
\providecommand \bibitemNoStop [0]{.\EOS\space}%
\providecommand \EOS [0]{\spacefactor3000\relax}%
\providecommand \BibitemShut  [1]{\csname bibitem#1\endcsname}%
\let\auto@bib@innerbib\@empty
\bibitem [{\citenamefont {Bhattacharjee}\ and\ \citenamefont {Dutta}(2021)}]{Bhattacharjee2021}%
  \BibitemOpen
  \bibfield  {author} {\bibinfo {author} {\bibfnamefont {S.}~\bibnamefont {Bhattacharjee}}\ and\ \bibinfo {author} {\bibfnamefont {A.}~\bibnamefont {Dutta}},\ }\href {https://doi.org/10.1140/epjb/s10051-021-00235-3} {\bibinfo {title} {Quantum thermal machines and batteries}} (\bibinfo {year} {2021})\BibitemShut {NoStop}%
\bibitem [{\citenamefont {Huang}\ \emph {et~al.}(2012)\citenamefont {Huang}, \citenamefont {Wang},\ and\ \citenamefont {Yi}}]{Huang2012}%
  \BibitemOpen
  \bibfield  {author} {\bibinfo {author} {\bibfnamefont {X.~L.}\ \bibnamefont {Huang}}, \bibinfo {author} {\bibfnamefont {T.}~\bibnamefont {Wang}},\ and\ \bibinfo {author} {\bibfnamefont {X.~X.}\ \bibnamefont {Yi}},\ }\href {https://doi.org/10.1103/PhysRevE.86.051105} {\bibfield  {journal} {\bibinfo  {journal} {Phys. Rev. E}\ }\textbf {\bibinfo {volume} {86}},\ \bibinfo {pages} {051105} (\bibinfo {year} {2012})}\BibitemShut {NoStop}%
\bibitem [{\citenamefont {Thomas}\ \emph {et~al.}(2018)\citenamefont {Thomas}, \citenamefont {Siddharth}, \citenamefont {Banerjee},\ and\ \citenamefont {Ghosh}}]{Thomas2018}%
  \BibitemOpen
  \bibfield  {author} {\bibinfo {author} {\bibfnamefont {G.}~\bibnamefont {Thomas}}, \bibinfo {author} {\bibfnamefont {N.}~\bibnamefont {Siddharth}}, \bibinfo {author} {\bibfnamefont {S.}~\bibnamefont {Banerjee}},\ and\ \bibinfo {author} {\bibfnamefont {S.}~\bibnamefont {Ghosh}},\ }\href {https://doi.org/10.1103/PhysRevE.97.062108} {\bibfield  {journal} {\bibinfo  {journal} {Phys. Rev. E}\ }\textbf {\bibinfo {volume} {97}},\ \bibinfo {pages} {062108} (\bibinfo {year} {2018})}\BibitemShut {NoStop}%
\bibitem [{\citenamefont {Campo}\ \emph {et~al.}(2014)\citenamefont {Campo}, \citenamefont {Goold},\ and\ \citenamefont {Paternostro}}]{DelCampo2014}%
  \BibitemOpen
  \bibfield  {author} {\bibinfo {author} {\bibfnamefont {A.~D.}\ \bibnamefont {Campo}}, \bibinfo {author} {\bibfnamefont {J.}~\bibnamefont {Goold}},\ and\ \bibinfo {author} {\bibfnamefont {M.}~\bibnamefont {Paternostro}},\ }\href {https://doi.org/10.1038/srep06208} {\bibfield  {journal} {\bibinfo  {journal} {Sci. Rep.}\ }\textbf {\bibinfo {volume} {4}},\ \bibinfo {pages} {6208} (\bibinfo {year} {2014})}\BibitemShut {NoStop}%
\bibitem [{\citenamefont {Roßnagel}\ \emph {et~al.}(2014)\citenamefont {Roßnagel}, \citenamefont {Abah}, \citenamefont {Schmidt-Kaler}, \citenamefont {Singer},\ and\ \citenamefont {Lutz}}]{Rossnagel2014}%
  \BibitemOpen
  \bibfield  {author} {\bibinfo {author} {\bibfnamefont {J.}~\bibnamefont {Roßnagel}}, \bibinfo {author} {\bibfnamefont {O.}~\bibnamefont {Abah}}, \bibinfo {author} {\bibfnamefont {F.}~\bibnamefont {Schmidt-Kaler}}, \bibinfo {author} {\bibfnamefont {K.}~\bibnamefont {Singer}},\ and\ \bibinfo {author} {\bibfnamefont {E.}~\bibnamefont {Lutz}},\ }\href {https://doi.org/10.1103/PhysRevLett.112.030602} {\bibfield  {journal} {\bibinfo  {journal} {Phys. Rev. Lett.}\ }\textbf {\bibinfo {volume} {112}},\ \bibinfo {pages} {030602} (\bibinfo {year} {2014})}\BibitemShut {NoStop}%
\bibitem [{\citenamefont {Kosloff}\ and\ \citenamefont {Rezek}(2017)}]{Kosloff2017}%
  \BibitemOpen
  \bibfield  {author} {\bibinfo {author} {\bibfnamefont {R.}~\bibnamefont {Kosloff}}\ and\ \bibinfo {author} {\bibfnamefont {Y.}~\bibnamefont {Rezek}},\ }\href {https://doi.org/10.3390/e19040136} {\bibinfo {title} {The quantum harmonic otto cycle}} (\bibinfo {year} {2017})\BibitemShut {NoStop}%
\bibitem [{\citenamefont {Cavaliere}\ \emph {et~al.}(2022)\citenamefont {Cavaliere}, \citenamefont {Carrega}, \citenamefont {Filippis}, \citenamefont {Cataudella}, \citenamefont {Benenti},\ and\ \citenamefont {Sassetti}}]{Cavaliere2022}%
  \BibitemOpen
  \bibfield  {author} {\bibinfo {author} {\bibfnamefont {F.}~\bibnamefont {Cavaliere}}, \bibinfo {author} {\bibfnamefont {M.}~\bibnamefont {Carrega}}, \bibinfo {author} {\bibfnamefont {G.~D.}\ \bibnamefont {Filippis}}, \bibinfo {author} {\bibfnamefont {V.}~\bibnamefont {Cataudella}}, \bibinfo {author} {\bibfnamefont {G.}~\bibnamefont {Benenti}},\ and\ \bibinfo {author} {\bibfnamefont {M.}~\bibnamefont {Sassetti}},\ }\href {https://doi.org/10.1103/physrevresearch.4.033233} {\bibfield  {journal} {\bibinfo  {journal} {Phys. Rev. Res.}\ }\textbf {\bibinfo {volume} {4}},\ \bibinfo {pages} {033233} (\bibinfo {year} {2022})}\BibitemShut {NoStop}%
\bibitem [{\citenamefont {Fei}\ \emph {et~al.}(2022)\citenamefont {Fei}, \citenamefont {Chen},\ and\ \citenamefont {Ma}}]{Fei2022}%
  \BibitemOpen
  \bibfield  {author} {\bibinfo {author} {\bibfnamefont {Z.}~\bibnamefont {Fei}}, \bibinfo {author} {\bibfnamefont {J.~F.}\ \bibnamefont {Chen}},\ and\ \bibinfo {author} {\bibfnamefont {Y.~H.}\ \bibnamefont {Ma}},\ }\href {https://doi.org/10.1103/PhysRevA.105.022609} {\bibfield  {journal} {\bibinfo  {journal} {Phys. Rev. A}\ }\textbf {\bibinfo {volume} {105}},\ \bibinfo {pages} {022609} (\bibinfo {year} {2022})}\BibitemShut {NoStop}%
\bibitem [{\citenamefont {Boubakour}\ \emph {et~al.}(2023)\citenamefont {Boubakour}, \citenamefont {Fogarty},\ and\ \citenamefont {Busch}}]{Boubakour2023}%
  \BibitemOpen
  \bibfield  {author} {\bibinfo {author} {\bibfnamefont {M.}~\bibnamefont {Boubakour}}, \bibinfo {author} {\bibfnamefont {T.}~\bibnamefont {Fogarty}},\ and\ \bibinfo {author} {\bibfnamefont {T.}~\bibnamefont {Busch}},\ }\href {https://doi.org/10.1103/PhysRevResearch.5.013088} {\bibfield  {journal} {\bibinfo  {journal} {Phys. Rev. Res.}\ }\textbf {\bibinfo {volume} {5}},\ \bibinfo {pages} {013088} (\bibinfo {year} {2023})}\BibitemShut {NoStop}%
\bibitem [{\citenamefont {Zhang}\ \emph {et~al.}(2014)\citenamefont {Zhang}, \citenamefont {Bariani},\ and\ \citenamefont {Meystre}}]{Zhang2014}%
  \BibitemOpen
  \bibfield  {author} {\bibinfo {author} {\bibfnamefont {K.}~\bibnamefont {Zhang}}, \bibinfo {author} {\bibfnamefont {F.}~\bibnamefont {Bariani}},\ and\ \bibinfo {author} {\bibfnamefont {P.}~\bibnamefont {Meystre}},\ }\href {https://doi.org/10.1103/PhysRevLett.112.150602} {\bibfield  {journal} {\bibinfo  {journal} {Phys. Rev. Lett.}\ }\textbf {\bibinfo {volume} {112}},\ \bibinfo {pages} {150602} (\bibinfo {year} {2014})}\BibitemShut {NoStop}%
\bibitem [{\citenamefont {Elouard}\ \emph {et~al.}(2017{\natexlab{a}})\citenamefont {Elouard}, \citenamefont {Herrera-Martí}, \citenamefont {Huard},\ and\ \citenamefont {Auffèves}}]{Elouard2017}%
  \BibitemOpen
  \bibfield  {author} {\bibinfo {author} {\bibfnamefont {C.}~\bibnamefont {Elouard}}, \bibinfo {author} {\bibfnamefont {D.}~\bibnamefont {Herrera-Martí}}, \bibinfo {author} {\bibfnamefont {B.}~\bibnamefont {Huard}},\ and\ \bibinfo {author} {\bibfnamefont {A.}~\bibnamefont {Auffèves}},\ }\href {https://doi.org/10.1103/PhysRevLett.118.260603} {\bibfield  {journal} {\bibinfo  {journal} {Phys. Rev. Lett.}\ }\textbf {\bibinfo {volume} {118}},\ \bibinfo {pages} {260603} (\bibinfo {year} {2017}{\natexlab{a}})}\BibitemShut {NoStop}%
\bibitem [{\citenamefont {Elouard}\ \emph {et~al.}(2017{\natexlab{b}})\citenamefont {Elouard}, \citenamefont {Herrera-Martí}, \citenamefont {Clusel},\ and\ \citenamefont {Auffèves}}]{Elouard2017a}%
  \BibitemOpen
  \bibfield  {author} {\bibinfo {author} {\bibfnamefont {C.}~\bibnamefont {Elouard}}, \bibinfo {author} {\bibfnamefont {D.~A.}\ \bibnamefont {Herrera-Martí}}, \bibinfo {author} {\bibfnamefont {M.}~\bibnamefont {Clusel}},\ and\ \bibinfo {author} {\bibfnamefont {A.}~\bibnamefont {Auffèves}},\ }\href {https://doi.org/10.1038/s41534-017-0008-4} {\bibfield  {journal} {\bibinfo  {journal} {Npj Quantum Inf.}\ }\textbf {\bibinfo {volume} {3}},\ \bibinfo {pages} {1} (\bibinfo {year} {2017}{\natexlab{b}})}\BibitemShut {NoStop}%
\bibitem [{\citenamefont {Elouard}\ and\ \citenamefont {Jordan}(2018)}]{Elouard2018}%
  \BibitemOpen
  \bibfield  {author} {\bibinfo {author} {\bibfnamefont {C.}~\bibnamefont {Elouard}}\ and\ \bibinfo {author} {\bibfnamefont {A.~N.}\ \bibnamefont {Jordan}},\ }\href {https://doi.org/10.1103/PhysRevLett.120.260601} {\bibfield  {journal} {\bibinfo  {journal} {Phys. Rev. Lett}\ }\textbf {\bibinfo {volume} {120}},\ \bibinfo {pages} {260601} (\bibinfo {year} {2018})}\BibitemShut {NoStop}%
\bibitem [{\citenamefont {Das}\ and\ \citenamefont {Ghosh}(2019)}]{Das2019}%
  \BibitemOpen
  \bibfield  {author} {\bibinfo {author} {\bibfnamefont {A.}~\bibnamefont {Das}}\ and\ \bibinfo {author} {\bibfnamefont {S.}~\bibnamefont {Ghosh}},\ }\href {https://doi.org/10.3390/e21111131} {\bibfield  {journal} {\bibinfo  {journal} {Entropy}\ }\textbf {\bibinfo {volume} {21}},\ \bibinfo {pages} {1131} (\bibinfo {year} {2019})}\BibitemShut {NoStop}%
\bibitem [{\citenamefont {Bresque}\ \emph {et~al.}(2021)\citenamefont {Bresque}, \citenamefont {Camati}, \citenamefont {Rogers}, \citenamefont {Murch}, \citenamefont {Jordan},\ and\ \citenamefont {Auffèves}}]{Bresque2021}%
  \BibitemOpen
  \bibfield  {author} {\bibinfo {author} {\bibfnamefont {L.}~\bibnamefont {Bresque}}, \bibinfo {author} {\bibfnamefont {P.~A.}\ \bibnamefont {Camati}}, \bibinfo {author} {\bibfnamefont {S.}~\bibnamefont {Rogers}}, \bibinfo {author} {\bibfnamefont {K.}~\bibnamefont {Murch}}, \bibinfo {author} {\bibfnamefont {A.~N.}\ \bibnamefont {Jordan}},\ and\ \bibinfo {author} {\bibfnamefont {A.}~\bibnamefont {Auffèves}},\ }\href {https://doi.org/10.1103/PhysRevLett.126.120605} {\bibfield  {journal} {\bibinfo  {journal} {Phys. Rev. Lett.}\ }\textbf {\bibinfo {volume} {126}},\ \bibinfo {pages} {120605} (\bibinfo {year} {2021})}\BibitemShut {NoStop}%
\bibitem [{\citenamefont {Manikandan}\ \emph {et~al.}(2022)\citenamefont {Manikandan}, \citenamefont {Elouard}, \citenamefont {Murch}, \citenamefont {Auffèves},\ and\ \citenamefont {Jordan}}]{Manikandan2022}%
  \BibitemOpen
  \bibfield  {author} {\bibinfo {author} {\bibfnamefont {S.~K.}\ \bibnamefont {Manikandan}}, \bibinfo {author} {\bibfnamefont {C.}~\bibnamefont {Elouard}}, \bibinfo {author} {\bibfnamefont {K.~W.}\ \bibnamefont {Murch}}, \bibinfo {author} {\bibfnamefont {A.}~\bibnamefont {Auffèves}},\ and\ \bibinfo {author} {\bibfnamefont {A.~N.}\ \bibnamefont {Jordan}},\ }\href {https://doi.org/10.1103/PhysRevE.105.044137} {\bibfield  {journal} {\bibinfo  {journal} {Phys. Rev. E}\ }\textbf {\bibinfo {volume} {105}},\ \bibinfo {pages} {044137} (\bibinfo {year} {2022})}\BibitemShut {NoStop}%
\bibitem [{\citenamefont {Étienne Jussiau}\ \emph {et~al.}(2023)\citenamefont {Étienne Jussiau}, \citenamefont {Bresque}, \citenamefont {Auffèves}, \citenamefont {Murch},\ and\ \citenamefont {Jordan}}]{Jussiau2023}%
  \BibitemOpen
  \bibfield  {author} {\bibinfo {author} {\bibnamefont {Étienne Jussiau}}, \bibinfo {author} {\bibfnamefont {L.}~\bibnamefont {Bresque}}, \bibinfo {author} {\bibfnamefont {A.}~\bibnamefont {Auffèves}}, \bibinfo {author} {\bibfnamefont {K.~W.}\ \bibnamefont {Murch}},\ and\ \bibinfo {author} {\bibfnamefont {A.~N.}\ \bibnamefont {Jordan}},\ }\href {https://doi.org/10.1103/PhysRevResearch.5.033122} {\bibfield  {journal} {\bibinfo  {journal} {Phys. Rev. Res.}\ }\textbf {\bibinfo {volume} {5}},\ \bibinfo {pages} {033122} (\bibinfo {year} {2023})}\BibitemShut {NoStop}%
\bibitem [{\citenamefont {Bezanson}\ \emph {et~al.}(2017)\citenamefont {Bezanson}, \citenamefont {Edelman}, \citenamefont {Karpinski},\ and\ \citenamefont {Shah}}]{Bezanson2017}%
  \BibitemOpen
  \bibfield  {author} {\bibinfo {author} {\bibfnamefont {J.}~\bibnamefont {Bezanson}}, \bibinfo {author} {\bibfnamefont {A.}~\bibnamefont {Edelman}}, \bibinfo {author} {\bibfnamefont {S.}~\bibnamefont {Karpinski}},\ and\ \bibinfo {author} {\bibfnamefont {V.~B.}\ \bibnamefont {Shah}},\ }\bibfield  {journal} {\bibinfo  {journal} {Society for Industrial and Applied Mathematics}\ }\textbf {\bibinfo {volume} {59}},\ \href {https://doi.org/10.1137/141000671} {10.1137/141000671} (\bibinfo {year} {2017})\BibitemShut {NoStop}%
\bibitem [{\citenamefont {Danisch}\ and\ \citenamefont {Krumbiegel}(2021)}]{Danisch2021}%
  \BibitemOpen
  \bibfield  {author} {\bibinfo {author} {\bibfnamefont {S.}~\bibnamefont {Danisch}}\ and\ \bibinfo {author} {\bibfnamefont {J.}~\bibnamefont {Krumbiegel}},\ }\href {https://doi.org/10.21105/joss.03349} {\bibfield  {journal} {\bibinfo  {journal} {Journal of Open Source Software}\ }\textbf {\bibinfo {volume} {6}},\ \bibinfo {pages} {3349} (\bibinfo {year} {2021})}\BibitemShut {NoStop}%
\bibitem [{\citenamefont {Wong}(2011)}]{Wong2011}%
  \BibitemOpen
  \bibfield  {author} {\bibinfo {author} {\bibfnamefont {B.}~\bibnamefont {Wong}},\ }\href {https://doi.org/10.1038/nmeth.1618} {\bibfield  {journal} {\bibinfo  {journal} {Nature Methods}\ }\textbf {\bibinfo {volume} {8}},\ \bibinfo {pages} {441} (\bibinfo {year} {2011})}\BibitemShut {NoStop}%
\bibitem [{\citenamefont {Gradshteyn}\ and\ \citenamefont {Ryzhik}(2015)}]{Gradshteyn}%
  \BibitemOpen
  \bibfield  {author} {\bibinfo {author} {\bibfnamefont {I.~S.}\ \bibnamefont {Gradshteyn}}\ and\ \bibinfo {author} {\bibfnamefont {I.~M.}\ \bibnamefont {Ryzhik}},\ }\href {https://doi.org/10.1016/C2010-0-64839-5} {\emph {\bibinfo {title} {Table of Integrals, Series, and Products: Eighth Edition}}},\ \bibinfo {edition} {8th}\ ed.,\ edited by\ \bibinfo {editor} {\bibfnamefont {D.}~\bibnamefont {Zwillinger}}\ and\ \bibinfo {editor} {\bibfnamefont {V.}~\bibnamefont {Moll}}\ (\bibinfo  {publisher} {Elsevier},\ \bibinfo {year} {2015})\BibitemShut {NoStop}%
\end{thebibliography}
%

\end{document}